\documentclass[11pt,twoside,english]{myamsart}
\advance\oddsidemargin by -1.0cm
\advance\evensidemargin by -1.0cm
\advance\topmargin by -1.0cm 
\usepackage{amssymb}
\usepackage{babel}
\usepackage{amstext}
\usepackage{amscd}   
\usepackage{epsfig}  
\usepackage{rotating}

\theoremstyle{plain}
\newtheorem{cor}{Corollary}[section]
\newtheorem{lem}{Lemma}[section]
\newtheorem{thm}{Theorem}[section]            

\theoremstyle{definition}
\newtheorem{exa}{Example}[section]
\newtheorem{NB}{Remark}[section]

\renewcommand{\labelenumi}{$(\theenumi)$}
\newcommand{\bdm}{\begin{displaymath}}
\newcommand{\edm}{\end{displaymath}}
\newcommand{\be}{\begin{equation}}
\newcommand{\ee}{\end{equation}}
\newcommand{\ba}[1]{\begin{array}{#1}}
\newcommand{\ea}{\end{array}}

\newcommand{\btab}{\begin{tabular}}
\newcommand{\etab}{\end{tabular}}

\newcommand{\R}{\ensuremath{\mathbb{R}}}

\newcommand{\D}{\ensuremath{\mathrm{D}}} 

\newcommand{\T}{\ensuremath{\mathrm{T}}}

\newcommand{\G}{\ensuremath{\mathrm{G}}}

\newcommand{\hut}{\wedge}

\newcommand{\Ric}{\ensuremath{\mathrm{Ric}}}
\newcommand{\Scal}{\ensuremath{\mathrm{Scal}}}

\newcommand{\U}{\ensuremath{\mathrm{U}}}

\newcommand{\SO}{\ensuremath{\mathrm{SO}}}

\begin{document}
\def\haken{\mathbin{\hbox to 6pt{%
                 \vrule height0.4pt width5pt depth0pt
                 \kern-.4pt
                 \vrule height6pt width0.4pt depth0pt\hss}}}
    \let \hook\intprod
\setcounter{equation}{0}
%
%
\thispagestyle{empty}
%
\date{\today}
\title[On the Ricci tensor in type II B string theory]
{On the Ricci tensor in type II B string theory}
%
%
%
\author{I. Agricola}
\author{T. Friedrich}
\author{P.-A. Nagy}
\author{C. Puhle}
\address{\hspace{-5mm} 
{\normalfont\ttfamily agricola@mathematik.hu-berlin.de}\newline
{\normalfont\ttfamily friedric@mathematik.hu-berlin.de}\newline
{\normalfont\ttfamily nagy@mathematik.hu-berlin.de}\newline
{\normalfont\ttfamily puhle@mathematik.hu-berlin.de}\newline
Institut f\"ur Mathematik \newline
Humboldt-Universit\"at zu Berlin\newline
Sitz: WBC Adlershof\newline
D-10099 Berlin, Germany}
%
\thanks{Supported by the SFB 647 "Raum, Zeit,Materie" and the SPP 1154 
``Globale Differentialgeometrie'' of the DFG as well as the
Volkswagen Foundation}
\subjclass[2000]{Primary 53 C 25; Secondary 81 T 30}
\begin{abstract}
Let $\nabla$ be a metric connection with totally skew-symmetric
torsion $\T$ on a Riemannian manifold. Given a spinor field $\Psi$ and
a dilaton function $\Phi$, the basic equations in type II B string
theory are
\bdm
\nabla \Psi \ = \ 0 \, , \quad \delta(\T) \ = \ a \cdot
\big(d \Phi
\haken \T \big) \, , \quad
\T \cdot \Psi \ = \ b \cdot d \Phi \cdot \Psi \, + \, 
\mu \cdot \Psi \, .
\edm
We derive some relations between the length $||\T||^2$ of the
torsion form, the scalar curvature of $\nabla$, the dilaton function
$\Phi$ and the parameters $a,b,\mu$. The main results deal with
the divergence of the Ricci tensor $\Ric^{\nabla}$ of the connection.
In particular, if the supersymmetry $\Psi$ is non-trivial and if
the conditions
\bdm
(d \Phi \haken \T) \haken \T \ = \ 0 \, , \quad
\delta^{\nabla}(d \T) \cdot \Psi = 0 
\edm
hold, then the energy-momentum tensor is 
divergence-free. We show that the latter condition is satisfied in many
examples constructed out of special geometries. A special case is $a = b$. 
Then the divergence of the energy-momentum tensor vanishes if and only if
one condition  $\delta^{\nabla}(d \T) \cdot \Psi = 0$ holds. 
Strong models ($d \T = 0$) 
have this property, but there are examples with $\delta^{\nabla}(d \T) 
\neq 0$ and $\delta^{\nabla}(d \T) \cdot \Psi = 0$.

\end{abstract}
\maketitle
\pagestyle{headings}
%
%
%
\section{Type II B string theory with constant dilaton}\noindent

\noindent
The mathematical model discussed in type II B string theory consists
of a Riemannian manifold $(M^n ,  g)$, a metric connection
$\nabla$ with totally skew-symmetric torsion $\T$ and a 
non-trivial spinor field $\Psi$. Putting the full Ricci tensor aside for
starters, there are three equations relating these objects:
\bdm\tag{$*$}
\nabla \Psi \ = \ 0 \, , \quad \delta(\T) \ = \ 0 \, , \quad
\T \cdot \Psi \ = \ \mu \cdot \Psi \, .
\edm
The spinor field describes the supersymmetry of the model. The first 
equation means that the spinor field $\Psi$ is parallel
with respect to the metric connection $\nabla$. The second equation
is a conservation law for the $3$-form $\T$. Since $\nabla$ is a metric
connection with totally skew-symmetric torsion, the divergences
$\delta^{\nabla}(\T) = \delta^g(\T)$ of the torsion form coincide
(see \cite{AgFr1}, \cite{FriedrichIvanov}). We denote this unique
$2$-form simply by $\delta(\T)$. The third equation is an
algebraic link between the torsion form $\T$ and the spinor field
$\Psi$. Indeed, the $3$-form $\T$ acts as an endomorphism in the spinor
bundle and the last equation requires that $\Psi$ is an eigenspinor for this
endomorphism.
There are models with $\mu = 0$ as well as models with a non-vanishing
$\mu$. In general, $\mu$ may be an arbitrary function. Since $\T$ acts 
on spinors as a symmetric endomorphism, $\mu$ has to be real. Moreover,
we will see that only real, constant parameters $\mu$ are possible.
It is well known (see \cite {FriedrichIvanov}) that the conservation
law $\delta(\T) = 0$ implies that the Ricci tensor $\Ric^{\nabla}$ of the
connection $\nabla$ is symmetric. Denote by $\Scal^{\nabla}$ the 
$\nabla$-scalar curvature and by $\Scal^g$ the scalar curvature of the
Riemannian metric. The existence of the $\nabla$-parallel spinor
field yields the so called integrability conditions (see \cite{Fri1}), 
i.\,e.~relations
between $\mu$, $\T$ and the curvature tensor of the connection $\nabla$. 
\begin{thm} \label{AlgebraIdent}
%
Let $(M^n,g,\nabla,\T,\Psi, \mu)$ be a solution of $(*)$ and assume that
the spinor field $\Psi$ is non-trivial. Then the function $\mu$ is constant
and we have 
\bdm
||\T||^2 \ = \ \mu^2 \, - \, \frac{\Scal^{\nabla}}{2} \ \geq \ 0 \, , \quad
\Scal^g \ = \ \frac{3}{2} \, \mu^2 \, + \, \frac{\Scal^{\nabla}}{4} \, . 
\edm
Moreover, the spinor field $\Psi$ is an eigenspinor of the endomorphism
defined by the $4$-form $d \T$,
\bdm
d \T \cdot \Psi \ = \ - \,\frac{\Scal^{\nabla}}{2} \cdot \Psi \, . 
\edm
\end{thm}
\begin{proof}
Let us associate 
with the $3$-form $\T$ the following $4$-form $\sigma_{\T}$, 
\bdm
\sigma_{\T}\ :=\ \frac{1}{2} \sum_{k=1}^n (e_k\haken\T)\hut
(e_k\haken\T) \, .
\edm
The square $\T^2$ of the $3$-form $\T$ in the Clifford algebra is 
the sum of a scalar and a $4$-form (see \cite{AgFr1}),
\bdm
\T^2 \, - \, |||\T||^2 \ = \ - \, 2 \cdot \sigma_{\T} \, .
\edm
The existence of a $\nabla$-parallel
spinor yields  the following condition (see \cite{FriedrichIvanov})
\bdm
3 \cdot d \T \cdot \Psi \, + \, 2 \cdot \delta(\T) \cdot \Psi \, - \, 
2 \cdot \sigma_{\T} \cdot \Psi \, + \, \Scal^{\nabla} \cdot \Psi \ = \ 0 \, .
\edm
Finally, there is a formula for the anti-commutator of the $\nabla$-Dirac
operator $\D_{\T}$ and the endomorphism $\T$ (see \cite{FriedrichIvanov}),
\bdm
\D_{\T} \circ \T \, + \, \T \circ \D_{\T} \ = \ 
d \T \, + \, \delta(\T) \, - \, 2 \cdot \sigma_{\T} \, - \, 
2 \, \sum_{i=1}^n (e_i \haken \T) \cdot \nabla_{e_i} \, . 
\edm
Combining these formulas we obtain, for example, 
\bdm
\mathrm{grad}(\mu) \cdot \Psi \ = \ \Scal^{\nabla} \cdot \Psi \, + \, 2 \cdot
(||\T||^2 \, - \, \mu^2) \cdot \Psi
\edm
and the result follows immediately.
\end{proof}
\noindent
Since $\mu$ has to be constant, equation 
$\T \cdot \Psi = \mu \cdot \Psi$ yields: 
\begin{cor}
For all vectors $X$,  one has
\bdm
(\nabla_X \T) \cdot \Psi \ = \ 0 \, .
\edm
\end{cor}
\begin{NB}
In particular,
the inequality $\Scal^{\nabla} \leq 2 \, \mu^2$ holds whenever there
exists a spinor field $\Psi\neq 0$ satisfying the equations. Moreover,
for a solution with  $\Scal^{\nabla} = 2\,\mu^2$,  the torsion form $\T$ 
has to vanish; this applies, in particular, to \emph{any} solution
of $(*)$ with $\mu=0$ (without any further assumption on the Ricci tensor
or the global structure of the underlying manifold!). 
This generalizes the observation (see \cite{Agri}) that 
the existence of a non-trivial solution of 
$\nabla \Psi = 0 \, , \, \Ric^{\nabla} = 0 \, , \,
\T \cdot \Psi = 0$ implies $\T = 0$ on \emph{compact} manifolds. 
It underlines the strength of the algebraic identities in 
Theorem~\ref{AlgebraIdent}.
\end{NB} 
\vspace{3mm}

\noindent
A further equation in type II B string theory deals with the Ricci
tensor $\Ric^{\nabla}$ of the connection. Usually one
requires for constant dilaton  that the
Ricci tensor has to vanish (see \cite{Gauntlett}). Understanding
this tensor as an energy-momentum tensor, it seems to be more
convenient to impose a weaker condition, namely
\bdm
\mathrm{div}(\Ric^{\nabla}) \ = \ 0 \, .
\edm
A subtle point is however the fact that there are a priori two different 
divergence operators. The first operator  $\mathrm{div}^g$
is defined by the Levi-Civita connection of the Riemannian metric, while
the second operator  $\mathrm{div}^{\nabla}$  is defined by the connection
$\nabla$.
It turns out that this difference does not play a role in
the formulation of the field equation under discussion. Moreover, under the 
assumption that a $\nabla$-parallel spinor exists, we can reformulate
the condition $\mathrm{div}(\Ric^{\nabla}) = 0$ in such a way that only
the spinor $\Psi$ and the torsion form $\T$ are involved. The next lemma,
although simple to prove, is crucial.
\begin{lem}\label{same-div}
 If $\nabla$ is a metric connection with totally skew-symmetric
torsion and $S$  a symmetric $2$-tensor, then 
\bdm
\mathrm{div}^g(S) \ = \ \mathrm{div}^{\nabla}(S) \, .
\edm
\end{lem}
\begin{proof}
The difference
\bdm
\mathrm{div}^g(S)(X) \, -  \, \mathrm{div}^{\nabla}(S)(X) \ = \ 
- \, \frac{1}{2} \sum_{i,j=1}^n S(e_i \, , \, e_j) \, 
\T(e_i \, , \, X \, , \, e_j) \ = \ 0
\edm
vanishes, since $S$ is symmetric and $\T$ is skew-symmetric.
\end{proof}
\begin{thm} \label{Ricci}
Let $(M^n,g,\nabla,\T,\Psi, \mu)$ be a solution of $(*)$,
\bdm
\nabla \Psi \ = \ 0 \, , \quad \delta(\T) \ = \ 0 \, , \quad
\T \cdot \Psi \ = \ \mu \cdot \Psi 
\edm
and assume that
the spinor field $\Psi$ is non-trivial. Then the Riemannian and the
$\nabla$-divergence of the Ricci tensor $\Ric^{\nabla}$ coincide,
$\mathrm{div}^g(\Ric^{\nabla}) = \mathrm{div}^{\nabla}(\Ric^{\nabla})$.
Moreover, $\mathrm{div}(\Ric^{\nabla})$ vanishes if and only if
$\delta^{\nabla}(d \T) \cdot \Psi = 0$ holds.
\end{thm}
\begin{proof}
The assumption $\delta(\T) = 0$ implies that the Ricci tensor
$\Ric^{\nabla}$ is symmetric (see \cite{FriedrichIvanov}). Therefore,
the vectors $\mathrm{div}^g(\Ric^{\nabla}) = 
\mathrm{div}^{\nabla}(\Ric^{\nabla})$ coincide by Lemma~\ref{same-div}.
Any $\nabla$-parallel spinor satisfies the condition (see 
\cite{FriedrichIvanov})
\bdm
\big( X \haken d \T \, + \, 2 \, \nabla_X \T \big) \cdot \Psi \, - \, 
2 \, \Ric^{\nabla}(X) \cdot \Psi \ = \ 0 \ .
\edm
Since we already know that $(\nabla_X\T)\cdot \Psi = 0$, the condition simplifies,
\bdm
\big( X \haken d \T \big) \cdot \Psi \, - \, 
2 \, \Ric^{\nabla}(X) \cdot \Psi \ = \ 0 \, .
\edm
First we differentiate this equation with respect to $\nabla$ and compute
the trace,
\bdm
\sum_{k=1}^n \nabla_{e_k} \big(e_k \haken d \T \big) \cdot \Psi \, - \, 
2\sum_{k=1}^n \nabla_{e_k} \big( \Ric^{\nabla}(e_k)\big) \cdot\Psi\ =\ 0\, .
\edm
The latter equation is equivalent to
\bdm
2 \, \mathrm{div}^g(\Ric^{\nabla}) \cdot \Psi \ = \
2 \, \mathrm{div}^{\nabla}(\Ric^{\nabla}) \cdot \Psi \ = \  
\delta^{\nabla}(d \T) \cdot \Psi \ . \qedhere
\edm
\end{proof}
\vspace{3mm}

\noindent
Tuples  $(M^n,g,\nabla,\T,\Psi, \mu)$ 
with a $\nabla$-parallel torsion form, $\nabla\T = 0$, are particularly 
interesting. This condition implies automatically the conservation 
law $\delta(\T)= 0$. Nearly
K\"ahler manifolds, Sasakian manifolds or nearly parallel $\G_2$-manifolds
in dimension $n=7$, all equipped with their unique characteristic
connection, are examples of metric connections with this property 
(see \cite{FriedrichIvanov}). 
In dimension $n=6$ we constructed several hermitian manifolds
of that type (\cite{AlFrSchoe}). 
Moreover, the canonical connection of any naturally reductive space
satisfies $\nabla\T = 0$ (see \cite{Agri}). The assumption $\nabla\T = 0$
implies that the length $||\T||^2$ is constant. If, moreover, 
there exists a spinor field $\Psi$ such that $\nabla \Psi =0 \, , \,
\T \cdot \Psi = \mu \cdot \Psi$, then by Theorem \ref{AlgebraIdent}
the scalar curvatures $\Scal^g$ and $\Scal^{\nabla}$ are constant.
On the other side, we use the formula
\bdm
0 \ = \ d^{\nabla} \T \ = \ \sum_{k=1}^n e_k \wedge \nabla_{e_k} \T \ = \ 
\sum_{k=1}^n e_k \wedge \nabla_{e_k}^g \T \, + \, \Sigma(\T \, , \, \T) \ = \ 
d \T \, + \, \Sigma(\T \, , \, \T) \, , 
\edm
where $\Sigma(\T \, , \, \T)$ is a quadratic expression in $\T$. 
Then we conclude that
\bdm
\nabla(d \T) \ = \ 0 \, , \quad \mbox{and} \quad 
\delta^{\nabla}( d \T) \ = \ 0 \ ,
\edm
i.\,e., we can apply Theorem \ref{Ricci}.
\begin{cor} \label{paralleleTorsion}
Let $(M^n,g,\nabla,\T,\Psi, \mu)$ be a tuple satisfying
\bdm
\nabla \Psi \ = \ 0 \, , \quad \nabla(\T) \ = \ 0 \, , \quad
\T \cdot \Psi \ = \ \mu \cdot \Psi 
\edm
and assume that
the spinor field $\Psi$ is non-trivial. Then the scalar curvatures are
constant and the divergence of the
Ricci tensor vanishes, $\mathrm{div}(\Ric^{\nabla}) = 0$.
\end{cor} 
%
\subsection{$5$-dimensional examples}\noindent
%
\vspace{3mm}

\noindent
Let $(M^5,g,\eta, \xi, \varphi)$ be a $5$-dimensional quasi-Sasakian
manifold. Its Nijenhuis tensor $\mathrm{N}$ vanishes and the 
fundamental form $\mathrm{F}$ is a closed $2$-form,
\bdm
\mathrm{N} \ = \ 0 \, , \quad d \mathrm{F} \ = \ 0 \, .
\edm
There exists a unique connection $\nabla$ preserving the contact structure
with totally skew-symmetric torsion, the \emph{characteristic connection}
of $(M^5,g,\eta, \xi, \varphi)$. Its torsion form is given by
(see \cite{FriedrichIvanov}, \cite{FriedrichIvanov2})
\bdm
\T \ = \ \eta \wedge d \eta \, .
\edm
If the differential $d \T = d \eta \wedge d \eta$ is proportional to 
$\mathrm{F} \wedge \mathrm{F}$ with a constant factor, the characteristic 
connection $\nabla$ (see \cite{Fri2} , \cite{AlFrSchoe})
of the $5$-manifold solves the equation
\bdm
\delta^{\nabla}(d \T) \ = \ 0 \, .
\edm
Indeed, $\nabla$ preserves the contact structure and we conclude 
that under this assumption the ``volume form''
$\mathrm{F} \wedge \mathrm{F}$ of the $4$-dimensional
bundle consisting of all vectors in $TM^5$ orthogonal to $\xi$ is
$\nabla$-parallel. In particular, $\delta^{\nabla}(d \T) = 0$ holds. 
In general, a quasi-Sasakian $5$-manifolds does not have to admit 
any $\nabla$-parallel spinor field. However, such examples are known and have 
been thoroughly investigated. Let us first consider the case 
of a Sasakian manifold, $d \eta =  2 \, \mathrm{F}$. There are Sasakian
$5$-manifolds admitting a $\nabla$-parallel spinor $\Psi$ such that
the following equations are satisfied (see \cite{FriedrichIvanov})
\bdm
\nabla \Psi \ = \ 0 \, , \quad \delta(\T) \ = \ 0 \, , \quad 
\T \cdot \Psi \ = \ \pm \, 4 \, \Psi \, , \quad
\mathrm{div}^g(\Ric^{\nabla}) \ = \ 0 \, .
\edm
The geometric data in these examples are
\bdm
||\T||^2 \ = \ 8 \, , \quad 
\Scal^{\nabla} \ = \ 16 \, , \quad \Scal^g \ = \ 28 \, . 
\edm
Moreover, there is a (locally) unique Sasakian $5$-manifolds admitting
a $\nabla$-parallel spinor field $\Psi$ such that $\T \cdot \Psi = 0$ holds
(see \cite{FriedrichIvanov2}).
It is the $5$-dimensional Heisenberg group equipped with its 
canonical Sasakian structure. In this case we have
\bdm
\nabla \Psi \ = \ 0 \, , \quad \delta(\T) \ = \ 0 \, , \quad 
\T \cdot \Psi \ = \ 0 \, , \quad
\mathrm{div}^g(\Ric^{\nabla}) \ = \ 0 
\edm
and the geometric data are
\bdm
||\T||^2 \ = \ 8 \, , \quad 
\Scal^{\nabla} \ = \ - \, 16 \, , \quad \Scal^g \ = \ - \, 4 \ .
\edm
In the paper \cite{FriedrichIvanov2}, we constructed a family
$M^5(a,b,c,d)$ depending on four real numbers $a,b,c,d$ of quasi-Sasakian
manifolds with $\nabla$-parallel spinor field $\Psi$. In this case we 
have
\bdm
\nabla \Psi \ = \ 0 \, , \quad \delta(\T) \ = \ 0 \, , \quad 
\T \cdot \Psi \ = \ \pm \, \sqrt{(a-d)^2 \, + \, 4 \, b^2 \, + \, 4 \, c^2}
\cdot \Psi \, , \quad
\mathrm{div}^g(\Ric^{\nabla}) \ = \ 0 
\edm
and the geometric data are
\bdm
||\T||^2 \ = \ a^2 \, + \, 2 \, b^2 \, + \, 2 \, c^2 \, + \, d^2 \, , \quad 
\Scal^{\nabla} \ = \ 4\, (b^2 \, + \, c^2 \, - \, a \, d) \, .
\edm
%
%
\subsection{$6$-dimensional examples}\noindent
%
\vspace{3mm}

\noindent
Let $(M^6, g , \mathrm{J})$ be a $6$-dimensional
nearly K\"ahler manifold. It admits a unique connection $\nabla$
with totally skew-symmetric torsion     
preserving the nearly K\"ahler structure  (see \cite{FriedrichIvanov},
\cite{AlFrSchoe}), which was first investigated  by A.~Gray
(\cite{Gray}). Moreover, there are two $\nabla$-parallel spinor
fields $\Psi^{\pm}$, and  there exists a positive number $a$ such that
\bdm
\nabla \Psi^{\pm} \ = \ 0 \, , \quad \delta(\T) \ = \ 0 \, , \quad 
\T \cdot \Psi^{\pm} \ = \ \pm \, 2 \, \sqrt{2} \, a \, \Psi^{\pm} \, , \quad
\mathrm{div}^g(\Ric^{\nabla}) \ = \ 0 \, .
\edm
The geometric data are
\bdm
||\T||^2 \ = \ 2 \, a \, , \quad 
\Scal^{\nabla} \ = \ 12 \, a \, , \quad \Scal^g \ = \ 15 \, a \, .
\edm
The Ricci tensors $\Ric^g$ and $\Ric^{\nabla}$ are proportional to the metric,
\bdm
\Ric^g \ = \ \frac{5}{2} \, a \,  \mathrm{Id} \, , \quad
\Ric^{\nabla} \ = \ 2 \, a \, \mathrm{Id} \, .
\edm
There is another interesting example. The paper \cite{AlFrSchoe} 
contains the construction of a hermitian $6$-manifold $(M^6,g,\mathrm{J})$
of type $\mathcal{W}_3$ such that its characteristic connection $\nabla$
has a $3$-dimensional, complex irreducible holonomy representation 
$\mathrm{Hol}(\nabla) \subset \U(3) \subset \SO(6)$. There
exist two $\nabla$-parallel spinor fields $\Psi^{\pm}$ and 
we have
\bdm
\nabla \Psi^{\pm} \ = \ 0 \, , \quad \delta(\T) \ = \ 0 \, , \quad 
\T \cdot \Psi^{\pm} \ = \ 0 \, , \quad
\mathrm{div}^g(\Ric^{\nabla}) \ = \ 0 .
\edm
The Ricci tensors are again proportional to the metric,
\bdm
\Ric^{\nabla} \ = \ - \, \frac{1}{3} \, ||\T||^2 \, \mathrm{Id} \, , \quad 
\Scal^{\nabla} \ = \ - \, 2 \, ||\T||^2 \, , 
\quad \Scal^g \ = \ - \, \frac{1}{2} \, ||\T||^2  .
\edm
%
%
\subsection{$7$-dimensional examples}\noindent
%
\vspace{3mm}

\noindent
Let $(M^7, g , \omega^3)$ be a 
$7$-dimensional nearly parallel $\G_2$-manifold. The equation 
$d \omega^3 = - \, a \,(* \omega^3)\, , \, a = \mathrm{constant} \neq 0$ 
characterizes this class of $\G_2$-manifolds. The torsion form 
of the characteristic connection is given by the formula (see
\cite{FriedrichIvanov})
\bdm
\T \ = \ - \, \frac{a}{6} \, \omega^3 .
\edm
There always exists a $\nabla$-parallel spinor field $\Psi$ and we have
\bdm
\nabla \Psi \ = \ 0 \, , \quad \delta(\T) \ = \ 0 \, , \quad 
\T \cdot \Psi \ = \  \frac{7}{6} \, a \, \Psi \, , \quad
\mathrm{div}^g(\Ric^{\nabla}) \ = \ 0 \, .
\edm
The Ricci tensors are again proportional to the metric (see 
\cite{FriedrichIvanov}, \cite{AgFr2}),
\bdm
\Ric^g \ = \ \frac{3}{8} \, a^2 \, \mathrm{Id} \, , \quad 
\Scal^g \ = \ \frac{21}{8} \, a^2 \, , \quad
\Scal^{\nabla} \ = \ \frac{7}{3} \, a^2 \, , \quad
||\T||^2 \ = \ \frac{7}{36} \, a^2 .
\edm
If the nearly parallel $\G_2$-structure is induced by an underlying
$3$-Sasakian structure,
we can construct a $2$-parameter family of torsion forms
$\T$ satisfying $\nabla\T = 0$ and admitting parallel spinors
(see \cite{AgFr1}).
Corollary \ref{paralleleTorsion} applies to this family, too.\\

\noindent
Let us next consider cocalibrated $\G_2$-manifolds 
such that the scalar product $(d \, \omega^3 , * \, \omega^3)$
is constant. $\G_2$-manifolds
of that type are characterized by the conditions (see \cite{Fernandez})
\bdm
d \, * \omega^3 \ = \ 0 \, , \quad (d \, \omega^3 \, , \, * \omega^3 ) \ = \ 
\mathrm{const} .
\edm
The torsion form $\T$ of its characteristic connection is given by the
formula (see \cite{FriedrichIvanov})
\bdm
\T \ = \ - \, * d \omega^3 \, + \, \frac{1}{6} \, 
(d \, \omega^3 \, , \, * \omega^3 ) \cdot \omega^3 .
\edm
There exists a $\nabla$-parallel spinor field $\Psi$, and for any
considered $\G_2$-manifold we have
\bdm
\nabla \Psi \ = \ 0 \, , \quad \delta(\T) \ = \ 0 \, , \quad
\T \cdot \Psi \ = \ - \, \frac{1}{6} \,
(d \, \omega^3 \, , \, * \omega^3 ) \, \Psi  \, .
\edm
The geometric data are given by (see \cite{FriedrichIvanov3}, \cite{AgFr2})
\bdm
\Scal^g \ = \ - \, \frac{1}{2} \, ||\T||^2  \, + \, \frac{1}{18} \, 
(d \, \omega^3 \, , \, * \omega^3 )^2  \, , \quad
\Scal^{\nabla} \ = \ - \, 2 \, ||\T||^2 \, + \, \frac{1}{18} \, 
(d \, \omega^3 \, , \, * \omega^3 )^2 . 
\edm
The Ricci tensor $\Ric^{\nabla}$ of the characteristic connection is
in general not divergence free, but  both
possible divergences coincide,
$\mathrm{div}^g(\Ric^{\nabla}) =  \mathrm{div}^{\nabla}(\Ric^{\nabla})$.
This vector is computable using the spinor field $\Psi$ and the
torsion form $\T$ only. On the other hand, a $3$-form $\pi^3$ vanishes
on the special parallel spinor $\Psi$ ($\pi^3 \cdot \Psi = 0$) if and
only if the $3$-form satisfies the following two algebraic equations
(see \cite{FriedrichIvanov})
\bdm
\pi^3 \wedge \omega^3 \ = \ 0 \, , \quad
\pi^3 \wedge (* \omega^3) \ = \ 0 \ .
\edm
This algebraic observation yields the following result.
\begin{cor}
Let $(M^7,g,\omega^3)$ be a cocalibrated $\G_2$-manifold 
such that the scalar product $ (d \, \omega^3 , * \omega^3 )$
is constant. Then the divergence of the Ricci tensor $\Ric^{\nabla}$ 
vanishes if and only if
\bdm
\delta^{\nabla}(d \T) \wedge \omega^3 \ = \ 0 \, , \quad
\delta^{\nabla}(d \T) \wedge (* \omega^3) \ = \ 0 \, .
\edm
\end{cor}
\begin{exa}
There exist $\G_2$-structures of pure type $\mathcal{W}_3$ in the
Fernandez/Gray classification (see \cite{Fernandez}) 
on the product of $\R^1$ by the $6$-dimensional
Heisenberg group and on the product of $\R^1$ by the $3$-dimensional
complex solvable Lie group. The torsion form of its characteristic connection
has been investigated in the paper \cite{FriedrichIvanov}. Using these
formulas, one computes directly that these examples satisfy
the conditions
$\delta^{\nabla}(d \T) \wedge \omega^3 =  0 \, , \,
\delta^{\nabla}(d \T) \wedge (* \omega^3) = 0$, but $\delta^{\nabla}(d \T)
  \neq 0$.
\end{exa}
%
\section{Type II B string theory with a dilaton function}\noindent
%

\noindent
In the first part of the paper, we discussed the Ricci tensor in the
model of type II B string theory. In fact, the model is much more flexible,
it contains an additional function $\Phi$. 
In the second part of the paper, we study the
corresponding results in this more general situation. 
We use basically the same arguments (although computationally more involved) 
as in the proofs of Theorems \ref{AlgebraIdent} and \ref{Ricci}, hence we 
shall not repeat them all. 
Again, the integrability conditions following from $\nabla\Psi = 0$ as
derived in \cite{FriedrichIvanov}, \cite{FriedrichIvanov3} are the
key ingredient.\\
\vspace{3mm}

\noindent
The equations now read as
\bdm\tag{$**$}
\nabla \Psi \ = \ 0 \, , \quad \delta(\T) \ = \ a \cdot
\big(d \Phi
\haken \T \big) \, , \quad
\T \cdot \Psi \ = \ b \cdot d \Phi \cdot \Psi \, + \, 
\mu \cdot \Psi \, .
\edm
Usually the constant $a$ has a precise value, namely
$a = \pm \, 2$.
In order to understand the role of the parameters in the equations, we slightly
generalized them and allow for two arbitrary parameters $a,b$. 
\begin{thm}
Let $(M^n,g,\nabla,\T,\Psi, \Phi, \mu, a)$ be a tuple satisfying $(**)$ 
and assume that the spinor field $\Psi$ is non-trivial. Then
\begin{gather}
(b \, - \, a) \cdot \delta(\T) \cdot \Psi = 0 \, , \quad
d \T \cdot \Psi \ = \ - \, \frac{\Scal^{\nabla}}{2} \cdot \Psi \, - \, 
\frac{b}{2} \, \Delta(\Phi) \cdot \Psi \, , \notag\\ 
||\T||^2 = \mu^2 \, - \, \frac{\Scal^{\nabla}}{2} \, + \, 
b^2 \, ||d \Phi||^2 \, - \, \frac{3b}{2} \, 
\Delta(\Phi) \, , \notag
\end{gather}
and  the Riemannian scalar curvature is given by the formula
\bdm
\Scal^g \ = \ \frac{3}{2} \, \mu^2 \, + \, \frac{3b^2}{2} \, 
||d \Phi||^2
\, + \, \frac{\Scal^{\nabla}}{4} \, - \, \frac{9\, b}{4} \, \Delta(\Phi) \, .
\edm
In particular, if $b \neq a$, we obtain $\delta(\T) \cdot \Psi = 0$. In this
case, the endomorphism $\T^2$ acts on the spinor by scalar multiplication,
\bdm
\T^2 \Psi \ = \ \big(b^2 \, ||d \Phi||^2 \, + \, \mu^2 \big)\cdot \Psi \, .
\edm
\end{thm}
\vspace{3mm}

\noindent
The differential $d \Phi$ of the dilaton  $\Phi$ is a $1$-form. Its
differentials $\nabla^g d \Phi \, , \, \nabla d \Phi$ \
with respect to the Levi-Civita connection $\nabla^g$ and
with respect to the connection $\nabla$, respectively, are bilinear forms.
Since the Levi-Civita connection is torsion-free, $\nabla^g d \Phi$ is
symmetric, $\nabla^g d \Phi(X ,  Y) = \nabla^g d \Phi(Y , X)$.
By Lemma~\ref{same-div}, one has
\bdm
\mathrm{div}^g(\nabla^g d \Phi)\  = \ \mathrm{div}^{\nabla}(\nabla^g
d\Phi) \, .
\edm  
The difference between the two bilinear forms is given by the torsion,
\bdm
\nabla d \Phi \ = \ \nabla^g d \Phi \, - \, \frac{1}{2} \cdot
(d \Phi \haken \T) \, .
\edm
Now we generalize Theorem \ref{Ricci}.
\begin{thm}
Let $(M^n,g,\nabla,\T,\Psi, \Phi, a, b, \mu)$ be a tuple satisfying $(**)$.
Then we have
\begin{eqnarray*}
2 \, \mathrm{div}^g (\Ric^{\nabla} \, - \, b \cdot \nabla^g d \Phi) \cdot \Psi 
&=& \delta^{\nabla} (d \T) \cdot \Psi \, + \, (a \, - \, b) \cdot
\delta^{\nabla}(d \Phi \haken \T) \cdot \Psi \ , \\
2 \, \mathrm{div}^{\nabla} (\Ric^{\nabla} \, - \, b \cdot \nabla^g d \Phi) 
\cdot \Psi 
&=& \delta^{\nabla} (d \T) \cdot \Psi \, - \, b \cdot
\delta^{\nabla}(d \Phi \haken \T) \cdot \Psi \ , \\
2 \, \mathrm{div}^g (\Ric^{\nabla} \, - \, b \cdot \nabla d \Phi) \cdot \Psi 
&=& \delta^{\nabla} (d \T) \cdot \Psi \, + \, (a \, - \, b) \cdot
\delta^{\nabla}(d \Phi \haken \T) \cdot \Psi \ , \\
2 \, \mathrm{div}^{\nabla} (\Ric^{\nabla} \, - \, b \cdot \nabla d \Phi) 
\cdot \Psi 
&=& \delta^{\nabla} (d \T) \cdot \Psi \, .
\end{eqnarray*}
In particular, the differences are given by ($\lambda\in\R$ is an arbitrary 
parameter)
\begin{eqnarray*}
2 \, (\mathrm{div}^g \, - \, \mathrm{div}^{\nabla}) 
(\Ric^{\nabla} \, - \, \lambda \cdot \nabla^g d \Phi)  
&=& a  \cdot \delta^{\nabla}(d \Phi \haken \T)  \ = \ 
\delta^{\nabla} \delta(\T)  \ , \\
2 \, (\mathrm{div}^g \, - \, \mathrm{div}^{\nabla}) 
(\Ric^{\nabla} \, - \, b \cdot \nabla d \Phi) 
&=& (a \, - \, b) \cdot \delta^{\nabla}(d \Phi \haken \T)  \, .
\end{eqnarray*}
\end{thm}
\begin{proof}
$2 \, \Ric^{\nabla} -  2 \, \lambda \, \nabla^g d\Phi +
\delta(\T)$ is a symmetric tensor. Hence, Lemma \ref{same-div} yields
\bdm
2 \, (\mathrm{div}^g \, - \, \mathrm{div}^{\nabla}) 
(\Ric^{\nabla} \, - \, \lambda \cdot \nabla^g d \Phi) \ = \ 
\delta^{\nabla}\delta(\T) \, - \, \delta^g \delta^g(\T) \ = \
\delta^{\nabla}\delta(\T) \, .
\edm
$2 \, \Ric^{\nabla} +  \delta(\T) -  2 \, b \, 
\nabla d \Phi -  b \, (d \Phi \haken \T)$ is symmetric, too. Consequently,
\begin{eqnarray*}
2 \, (\mathrm{div}^g \, - \, \mathrm{div}^{\nabla}) 
(\Ric^{\nabla} \, - \, b \cdot \nabla d \Phi) &=& 
\delta^{\nabla} \delta(\T) \, - \, b \, \delta^{\nabla}(d\Phi \haken \T)
\, - \, \delta^g \delta^g(\T) \\ 
&+&  b \, \delta^g(d \Phi \haken T) 
=\ (a \, - \, b) \cdot \delta^{\nabla}(d \Phi \haken \T) \, .
\end{eqnarray*}
Here we used once again the equation $\delta(\T) = a \cdot(d \Phi \haken \T)$.
The equation $\T \cdot \Psi = b \cdot d \Phi \cdot \Psi \, + \, \mu 
\cdot \Psi$ yields
\bdm
(\nabla_X\T) \cdot \Psi \ = \ b \cdot (\nabla_X d \Phi) \cdot \Psi \, .
\edm
Next we differentiate the integrability condition
\bdm
\big( X \haken d \T \, + \, 2 \, \nabla_X \T \big) \cdot \Psi \, - \, 
2 \, \Ric^{\nabla}(X) \cdot \Psi \ = \ 0 \ .
\edm
and proceed as in the proof of Theorem \ref{Ricci}. The result is a similar
one,
\bdm
2 \, \mathrm{div}^{\nabla} (\Ric^{\nabla} \, - \, b \cdot \nabla d \Phi) 
\cdot \Psi 
\ = \ \delta^{\nabla} (d \T) \cdot \Psi \, .
\edm
Finally, we have 
\begin{eqnarray*}
2 \, \mathrm{div}^{\nabla} (\Ric^{\nabla} \, - \, b \cdot \nabla^g d \Phi) 
\cdot \Psi &=& 2 \, \mathrm{div}^{\nabla} (\Ric^{\nabla} \, - \, b \cdot 
\nabla d \Phi) \cdot \Psi \\
&& + \, 2 \, b \, 
\mathrm{div}^{\nabla}(\nabla d \Phi \, - \, \nabla^g d \Phi) \cdot \Psi \\\
&=& \delta^{\nabla}(d\T) \cdot \Psi \, - \, b \cdot \delta^{\nabla}(d \Phi 
\haken \T) \cdot \Psi\, . 
\end{eqnarray*}
The remaining formulas now follow directly from what has already been shown.
\end{proof}
\vspace{3mm}

\noindent
Using the equation $\delta(\T) = a \cdot ( d\Phi \haken \T)$,
the formula $\delta^g \delta(\T) = \delta^g \delta^g(\T) = 0$ as well as
the formulas comparing $\delta^g$ and $\delta^{\nabla}$ on differential
forms (see \cite{AgFr1}), we compute that the $1$-form $\delta^{\nabla} 
\delta(\T)$ is proportional to the $1$-form $(d \Phi \haken \T) \haken \T$.
Consequently, we obtain a necessary and sufficient 
algebraic condition under which the different
divergences coincide.
\begin{cor}
If, in addition, $(d \Phi \haken \T) \haken \T = 0$ 
holds, then the following divergences coincide:
\begin{eqnarray*}
\mathrm{div}^g (\Ric^{\nabla} \, - \, \lambda \cdot \nabla^g d \Phi) &=& 
\mathrm{div}^{\nabla}(\Ric^{\nabla} \, - \, \lambda \cdot \nabla^g d \Phi)
\quad \mbox{for any} \ \ \lambda \in \R, \\
\mathrm{div}^g (\Ric^{\nabla} \, - \, b \cdot \nabla d \Phi) &=& 
\mathrm{div}^{\nabla}(\Ric^{\nabla} \, - \, b \cdot \nabla d \Phi) \ .
\end{eqnarray*}
\end{cor}
\begin{cor}
If $(d \Phi \haken \T) \haken \T = 0$ and $\delta^{\nabla}(d \T)
\cdot \Psi = 0$ hold, then all the divergences vanish.
\end{cor}
\vspace{3mm}

\noindent
The previous discussion shows that the case of $a = b$ is a special one.
Normalizing the constants, we assume that $a = b = - \, 2$. Then the equations
($**$) read as
\bdm
\nabla \Psi \ = \ 0 \, , \quad \delta(\T) \ = \ - \, 2 \cdot
\big(d \Phi
\haken \T \big) \, , \quad
\T \cdot \Psi \ = \ - \, 2 \cdot d \Phi \cdot \Psi \, + \, 
\mu \cdot \Psi \, .
\edm
In this case, the condition $\delta(\T) \cdot \Psi =0$ is
{\it not} an integrability condition and the correct formula for 
$\T^2 \cdot \Psi$ is different,
\bdm
\T^2 \cdot \Psi \ = \ ( 4 \, ||d \Phi||^2 \, + \, \mu^2) \cdot \Psi 
\, - \, 2 \cdot \delta(\T) \cdot \Psi \, .
\edm
The divergence of the energy-momentum tensor is given by
\bdm
2 \, \mathrm{div}^g (\Ric^{\nabla} \, + \, 2 \cdot \nabla^g d \Phi) \cdot \Psi 
\ = \ \delta^{\nabla} (d \T) \cdot \Psi \, .
\edm
In strong models ($d \T=0$) the divergence of the energy-momentum tensor
vanishes, but there are models with $\delta^{\nabla}(d \T) \neq 0$ and $
\delta^{\nabla} (d \T) \cdot \Psi =0$. The last example
in Section 1.3 is one of them.
    
\end{document}